\documentclass[fleqn,twoside]{article}
\usepackage{espcrc2}

\usepackage{graphicx}
\usepackage[figuresright]{rotating}
\usepackage{longtable}

\newcommand{\be}{\begin{equation}}
\newcommand{\ee}{\end{equation}}
\newcommand{\bea}{\begin{eqnarray}}
\newcommand{\eea}{\end{eqnarray}}

\newcommand{\wh}{\widehat}
\newcommand{\wt}{\widetilde}

\title{Determination of $\alpha_s(M_\tau^2)$:
a conformal mapping approach}

\author{Irinel Caprini\thanks{Speaker}\address{National Institute of Physics and Nuclear Engineering,  Bucharest POB MG-6, R-077125 Romania} and 
Jan Fischer\address{Institute of Physics, Academy of Sciences of the Czech Republic, 
CZ-182 21  Prague 8, Czech Republic}}
       
\begin{document}

\begin{abstract}
 We discuss a new class of  expansions in perturbative QCD, based on the technique of conformal mappings of the Borel plane, and apply them for the determination  of $\alpha_s$ from the hadronic decays of the $\tau$ lepton. Using the expansion up to fifth order in the $\overline{\rm MS}$ scheme, the method leads to the prediction  $\alpha_s(M_\tau^2)=  0.320\pm 0.011$.
\vspace{1pc}
\end{abstract}
\maketitle
 
\section{INTRODUCTION}\label{sec:intro}

The determination of the strong coupling  $\alpha_s$ is one of the most important tests of QCD.  The recent determinations at various scales are in an impressive agreement among each other, leading to the world average \cite{Bethke}
\be \label{eq:world2009}
\alpha_s(M_Z^2)=0.1184 \pm 0.0007. \ee 
The hadronic decays of the $\tau$ lepton provide one of the most precise ways of extracting $\alpha_s$.
The recent calculation of the Adler function to four loops \cite{BCK08} triggered several new determinations of $\alpha_s(M_\tau^2)$. One may note that the average  reported in \cite{Bethke}
\be  \label{eq:alphatau}
\alpha_s(M_\tau^2)=0.330 \pm 0.014,
\ee
leads to the value $\alpha_s(M_Z^2)=0.1197 \pm 0.0016$, slightly higher than the global average (\ref{eq:world2009}).

The  discrepancy  between the so-called contour-improved perturbation theory (CIPT) \cite{DiPi}  and the more usual fixed-order expansions (FOPT)  appears to be the largest systematic theoretical uncertainty in the  determination of $\alpha_s(M_\tau^2)$. The recent works \cite{Davier2008}-\cite{Pich2010} show that this discrepancy  did not go away by adding the presently known higher-order terms in the expansion of the Adler function.   An attempt to understand the difference between CIPT and FOPT  was performed in \cite{CaFi2009}, based on previous studies \cite{CaFi}, which exploit the information about the high-order behaviour of the perturbation expansion, encoded in the singularities of the Adler function in the Borel plane. Essentially, the usual series in powers of the Borel variable is replaced by an improved expansion in powers of an "optimal" variable that maps the  Borel cut-plane onto a disc. In the present contribution we extend the investigation by  considering other conformal mappings, useful when the known nature of the dominant singularities of the Borel tranform is also incorporated. We illustrate the usefulness of the new expansions using some realistic models of the Adler function, and  apply the method for a new determination of $\alpha_s(M_\tau^2)$.

\section{ADLER FUNCTION}\label{sec:adler}
The determination of  $\alpha_s(M_\tau^2)$ is based on the the evaluation of the integral
\be\label{eq:delta0}
\delta^{(0)} =  \frac{1}{2\pi i}\, \oint\limits_{|s|=M_\tau^2}\, \frac{d s}{s}\, \omega(s)\, \wh D(s),\ee
where $ \omega(s)=1-2s/M_\tau^2 + 2(s/M_\tau^2)^3-(s/M_\tau^2)^4$ and the Adler function $\wh D(s)$ is expressed in perturbative QCD as
\be \label{eq:Dpert}
\wh D(s) = \sum\limits_{n\ge 1} [K_{n}+\kappa_n(-s/\mu^2)]\,  (a_s(\mu^2))^n,
\ee
where $a_s(\mu^2)=\alpha_s(\mu^2)/\pi$.  The first coefficients  $K_n$ calculated in the $\overline{\rm MS}$ scheme are $K_1=1$, $K_2= 1.64$, $K_3=6.37$ and $K_4=49.08$. For the next term the choices $K_5=283$ and  $K_5=275$ were made recently in \cite{BeJa} and \cite{Pich2010}. Finally, $\kappa_n(-s/\mu^2)$ depend on the coefficients of the renormalization-group (RG), and the renormalization scale $\mu^2$ is chosen as $\mu^2=-s$ in  CIPT and $\mu^2=M_\tau^2$  in FOPT.

 Expansions alternative to (\ref{eq:Dpert}) were proposed in the literature from various motivations. In the present work, the objective is to include theoretical knowledge about the high order behaviour of the series. From particular classes of Feynman diagrams it is known that $K_n\sim n!$, so the renormalized perturbation series (\ref{eq:Dpert}) is divergent and is usually assumed to be an asymptotic series. From independent arguments it is known that correlation functions like $\wh D$, regarded as functions of $\alpha_s$, are singular at $\alpha_s=0$. For QED these facts are known since a long time \cite{Dyson}, but they do not affect the phenomenological predictions since the coupling is very small. By contrast,  for a large coupling like $\alpha_s(M_\tau^2)$ in  QCD  the consequences are nontrivial.

The information about the high-order behaviour of the series (\ref{eq:Dpert}) is included in the singularities of the Borel transform $B(u)$, defined by the series
\be\label{B} B(u)=\sum_{n\ge 0} b_n u^n,\quad \quad b_n=\frac{K_{n+1}}{\beta_0^n \,n!}\,,\quad n\ge 0, \ee
where $\beta_0$ is the first coefficient of the RG $\beta$-function. According to the present knowledge,  $B(u)$ has singularities on the real axis  for $u \le -1$ and $u\ge 2$, known as  ultraviolet (UV) and infrared (IR)  renormalons, respectively. 

The expansion (\ref{eq:Dpert}) can be formally obtained from $B(u)$ by means of an integral of Borel-Laplace type. The recovery of the function $\wh D(s)$ is actually ambiguous: there are many integral representations admitting (\ref{eq:Dpert}) as an asymptotic expansion (for a recent discussion see \cite{CaFiVr}). As shown in \cite{CaNe}, the principal value prescription
\be\label{eq:pv}
\wh D(s)=\frac{1}{\beta_0}\,{\rm PV} \,\int\limits_0^\infty  e^{-\frac{u}{\beta_0 a_s(s)}} \, B(u)\, {\rm d} u
\ee
yields a function $\wh D(s)$ satisfying to a large extent the general analyticity requirements in the $s$-plane, and we shall adopt here this definition. 

\section{CONVERGENCE ACCELERATION BY CONFORMAL MAPPINGS}\label{sec:confmap}

Due to the first UV renormalon at $u=-1$, the series (\ref{B}) converges only in the disc $|u|<1$. The domain of convergence of an expansion and its convergence rate can be increased by expanding the function in powers of a different variable.  The following two lemmas \cite{CiFi,CaFinew} show that one can find an optimal variable by means a conformal mapping of the $u$-plane.

\vspace{0.2cm}\noindent{\it Lemma 1}:~  Let ${\cal D}_1$  and  ${\cal D}_2$  be two simply-connected  domains in the complex $u$-plane, with
$ {\cal D}_2 \subset {\cal D}_1$. Consider the conformal mappings
$ z_1=\tilde z_1(u): {\cal D}_1\to {\cal D}=\{z_1:|z_1|<1\}$ and
$ z_2=\tilde z_2(u): {\cal D}_2\to {\cal D}=\{z_2:|z_2|<1\}$ 
such that $ z_1(0)=0$ and $ z_2(0)=0$. 
Then $ |\tilde z_1(u)| \le  |\tilde z_2(u)|$ for all $ u\in {\cal D}_2$.

\vspace{0.2cm}\noindent
{\it Lemma 2}:  Let ${\cal D}_1$  and  ${\cal D}_2$ be the domains defined in Lemma 1, and
 $B(u)$ holomorphic in ${\cal D}_1$. Define the expansions
$ B(u)=\sum_{0}^{\infty} c_{n,1} (\tilde z_1(u))^n$ and
 $B(u)=\sum_{0}^{\infty} c_{n,2} (\tilde z_2(u))^n$, 
 convergent in the unit discs $|z_1|<1$ and $|z_2|<1$, respectively. Then, at large $n$ one has
\be\label{eq:rate}
\left\vert\frac{c_{n,1} (\tilde z_1(u))^n}{c_{n,2} (\tilde z_2(u))^n}\right\vert <1.
\ee

\begin{table*}[!ht]
\caption{ The quantity $\delta^{(0)}$  for the model defined in \cite{BeJa}, calculated for  
$\alpha_s(M_\tau^2)=0.34$ with the standard and modified  CI and FO expansions truncated at the order $N$. Exact value $\delta^{(0)} =0.2371$.} \vspace{0.1cm}
\label{table:1}
\renewcommand{\tabcolsep}{0.5pc} 
\renewcommand{\arraystretch}{1.1} 
\begin{tabular}{llllllllllr}\hline
$N$&CIPT &FOPT& CI $w_{12}$  &  FO $w_{12}$ &CI $w_{13}$& FO $w_{13}$ & CI $w_{1\infty}$& FO $w_{1\infty}$& CI $w_{23}$ &FO $w_{23}$\\\hline 
2& 0.1776& 0.1692& 0.1977& 0.2228& 0.2070& 0.2203& 0.1883& 
0.2524& 0.2123& 0.2099\\
3& 0.1898& 0.2026& 0.2009& 0.2460& 
0.2030& 0.2440& 0.1975& 0.2530& 0.2028& 0.2437
\\ 4& 0.1983& 
0.2200& 0.2263& 0.2463& 0.2194& 0.2460& 0.2288& 0.2465& 0.2206& 
0.2463\\ 5& 0.2022& 0.2288 & 0.2290 & 0.2440 & 0.2268 & 0.2423 & 
0.2310 & 0.2427 & 0.2292 & 0.2423 \\ 6& 0.2046 & 0.2328 & 0.2324 & 
0.2484 & 0.2306 & 0.2421 & 0.2321 & 0.2431 & 0.2319 & 0.2449 \\7& 
0.2046 & 0.2342 & 0.2339 & 0.2536 & 0.2331 & 0.2457 & 0.2333 & 
0.2454 & 0.2345 & 0.2502\\ 8& 0.2017 & 0.2353 & 0.2339  & 0.2505 & 
0.2343 & 0.2484 & 0.2341 & 0.2471 & 0.2347 & 0.2476 \\ 9 & 0.2004 & 
0.2367 & 0.2341 & 0.2431 & 0.2348 & 0.2457 & 0.2346 & 0.2465 & 
0.2347 & 0.2377 \\ 10 & 0.1842 & 0.2390 & 0.2351 & 0.2420 & 0.2348 & 
0.2394 & 0.2348 & 0.2436 & 0.2353 & 0.2337\\11 & 0.1962 & 0.2402 & 
0.2359 & 0.2406 & 0.2348 & 0.2352 & 0.2349 & 0.2399 & 0.2348 & 
0.2335\\12 & 0.1123 & 0.2436 & 0.2362 & 0.2298 & 0.2351 & 0.2349 & 
0.2349 & 0.2370 & 0.2374 & 0.2262\\ 13 & 0.2629 & 0.2408 & 0.2362 & 
0.2229 & 0.2355 & 0.2341 & 0.2349 & 0.2356 & 0.2348 & 0.2226\\ 14 & 
-0.2915 & 0.2575 & 0.2364 & 0.2242 & 0.2361 & 0.2303 & 0.2349 & 
0.2354 & 0.2395 & 0.2314\\ 15 & 1.1011 & 0.2170 & 0.2367 & 0.2173 & 
0.2366 & 0.2277 & 0.2350 & 0.2357 & 0.2356 & 0.2365\\ 16 & -3.362 & 
0.3818 & 0.2368 & 0.2102 & 0.2369 & 0.2305 & 0.2351 & 0.2360 & 
0.2343 & 0.2374\\17 & 9.5931 & -0.1881 & 0.2368 & 0.2176 & 0.2372 & 
0.2356 & 0.2352 & 0.2360 & 0.2533 & 0.2512\\ 18 & -31.52 & 2.144 & 
0.2368 & 0.2201 & 0.2373 & 0.2371 & 0.2354 & 0.2359 & 0.1926 & 
0.2665\\\hline \end{tabular}
\end{table*}
From this inequality it follows that the best asymptotic convergence rate is obtained with the variable that maps the whole holomorphy domain onto the unit disc.
For the Adler function, assuming that there are no other singularities except the cuts along the real axis for $u\le -1$ and $u\ge 2$, the "optimal" conformal mapping is \cite{CaFi}
\be\label{eq:w}
\wt w(u)=\frac{\sqrt{1+u}-\sqrt{1-u/2}}{\sqrt{1+u}+\sqrt{1-u/2}}.\ee
It follows that the expansion
\be\label{eq:Bw}  B(u)=\sum_{n\ge 0} d_n \,w^n, \ee
where $w= \wt w(u)$, converges in $|w|<1$, {\em i.e.} in the whole cut $u$-plane, and has the best asymptotic convergence rate, compared to other conformal mappings which map only
a part of the cut plane onto the unit disc.
This led us to define the new perturbative expansion \cite{CaFi}
\be\label{eq:DnewCI}
\wh D(s)=\sum\limits_{n\ge 0} d_n {\cal W}_n(s), 
\ee
\vspace{-0.6cm}
\be\label{eq:Wn}
{\cal W}_n(s)=\frac{1}{\beta_0}{\rm PV} \int\limits_0^\infty\! \,{\rm e}^{-\frac{u}{\beta_0 a_s(s)}}  (\wt w(u))^n\,{\rm d}u.
\ee
By construction, the series (\ref{eq:DnewCI}), when reexpanded in powers of $\alpha_s$, reproduces the coefficients $ K_n$   known from Feynman diagrams. Moreover, under certain condition, the expansion (\ref{eq:DnewCI}) converges in a domain of the $s$-plane \cite{CaFi}.  On the other hand, as shown in \cite{CaFi}, the expansion functions ${\cal W}_n$ are singular at $\alpha_s=0$, resembling the expanded function $\wh D$ itself.  

\begin{figure*}[!ht]\label{fig:fig1}\begin{center}\includegraphics[width=6.cm]{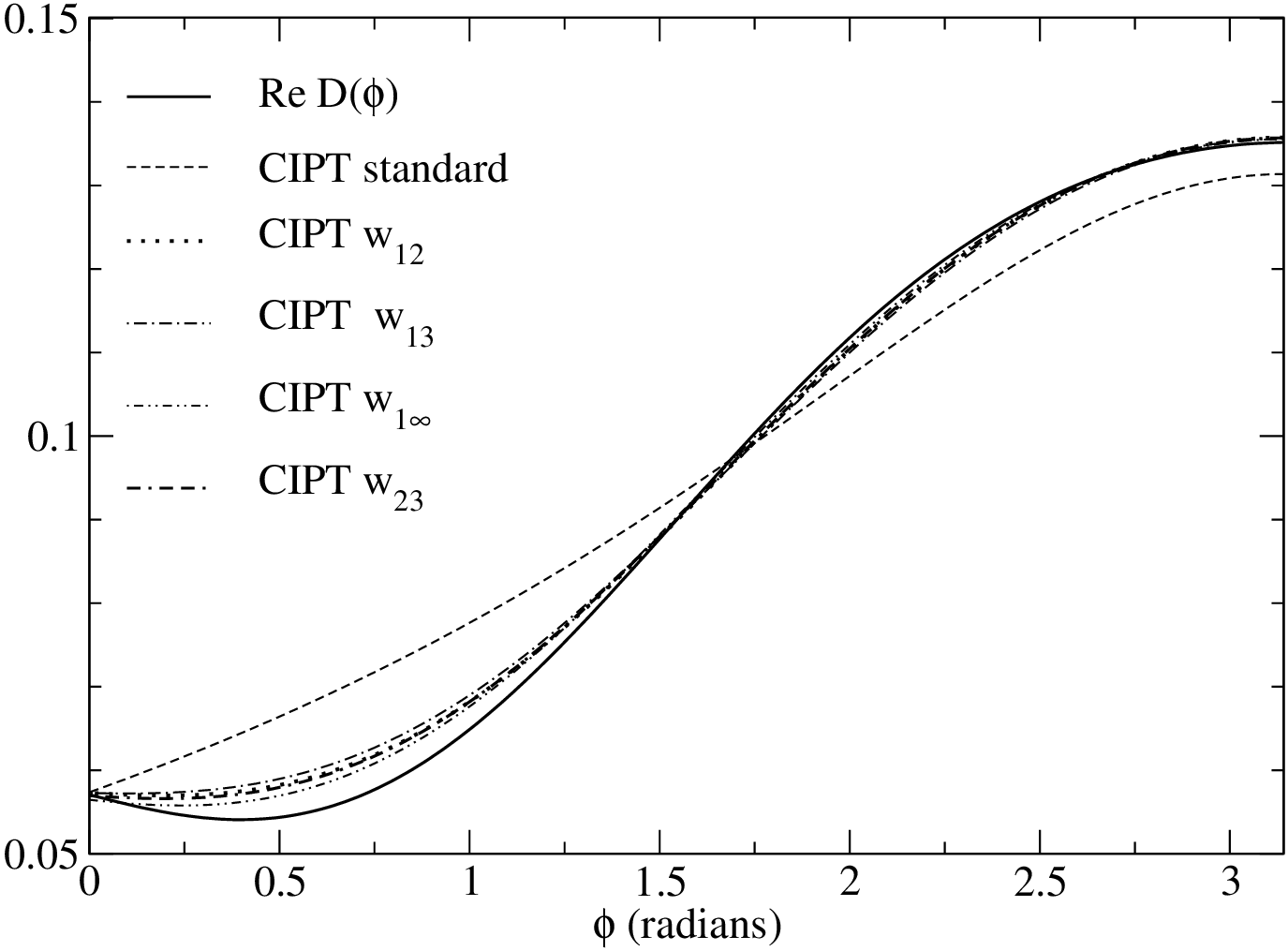}\hspace{1cm}
\includegraphics[width=6.cm]{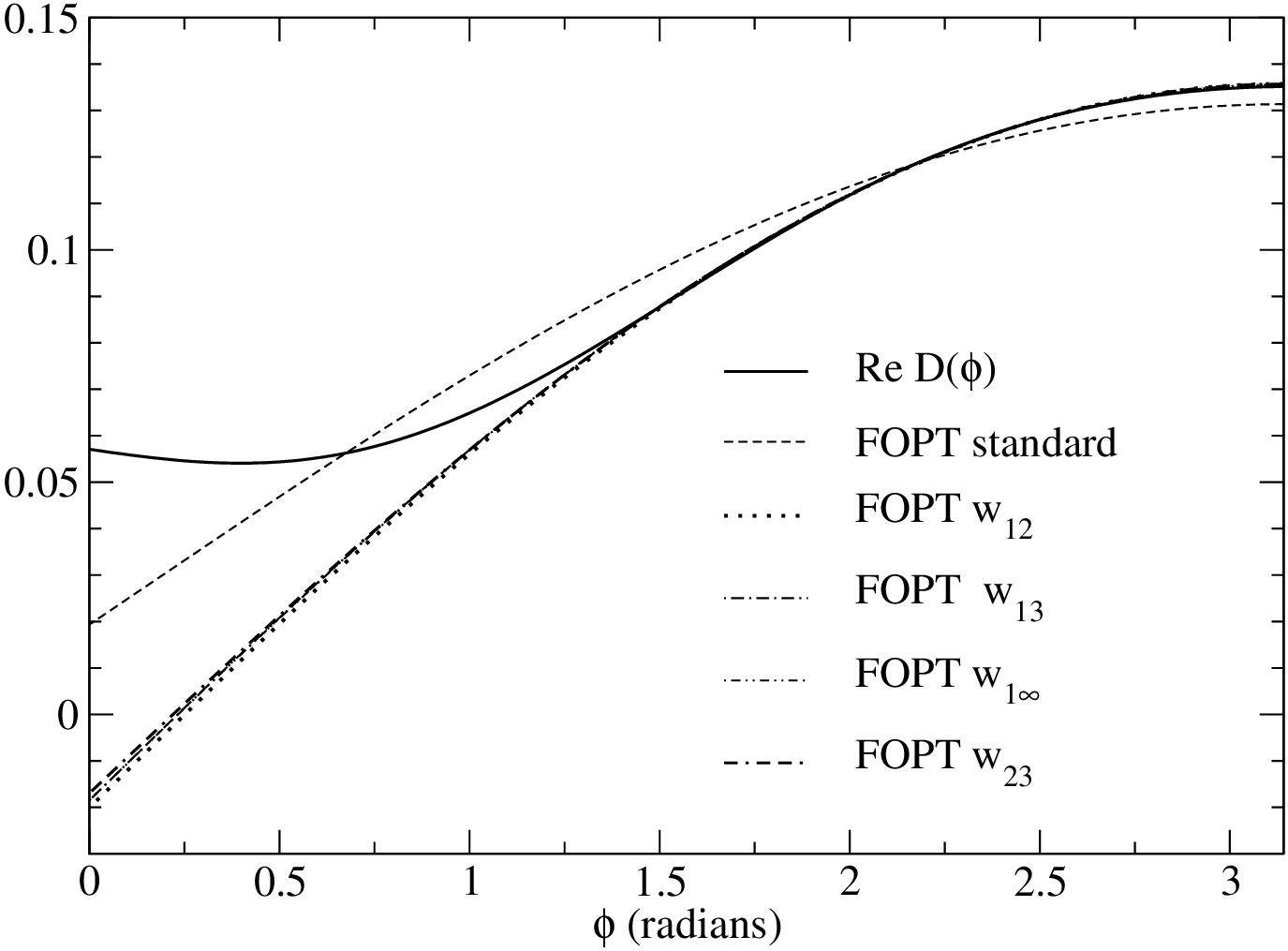}
\caption{Real part of the Adler function of the model \cite{BeJa}, calculated along the circle $s=M_\tau^2 \exp(i\phi)$ with the perturbative expansions for  $N=5$. Left panel: CI expansions. Right panel: FO expansions. The exact function is represented by the solid line. For details see \cite{CaFi2009,CaFinew}.}\end{center}\vspace{-0.3cm}
\end{figure*}
\begin{figure*}[!ht]\label{fig:fig2}\begin{center}\includegraphics[width=6.cm]{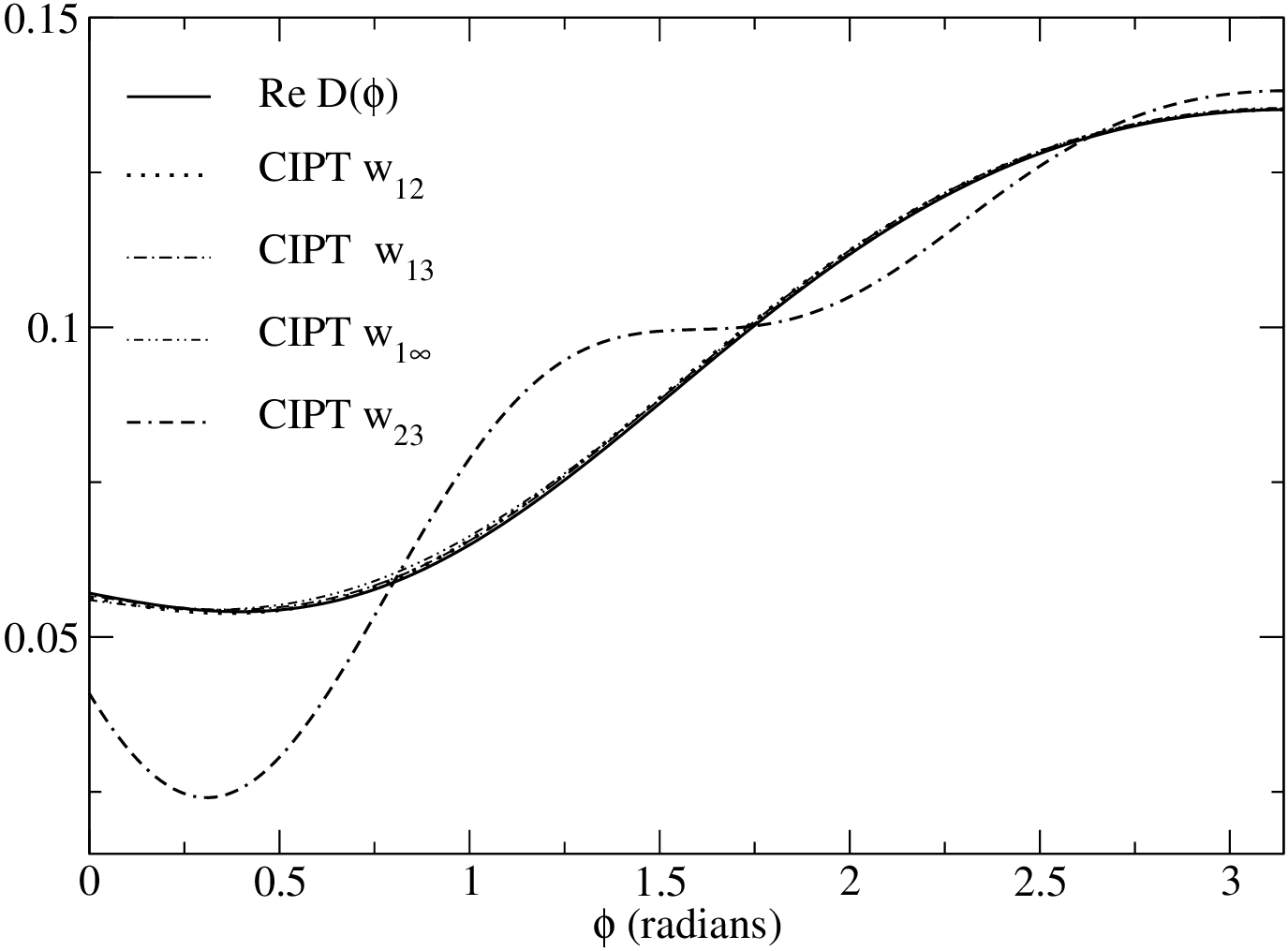}\hspace{1cm}\includegraphics[width=6.cm]{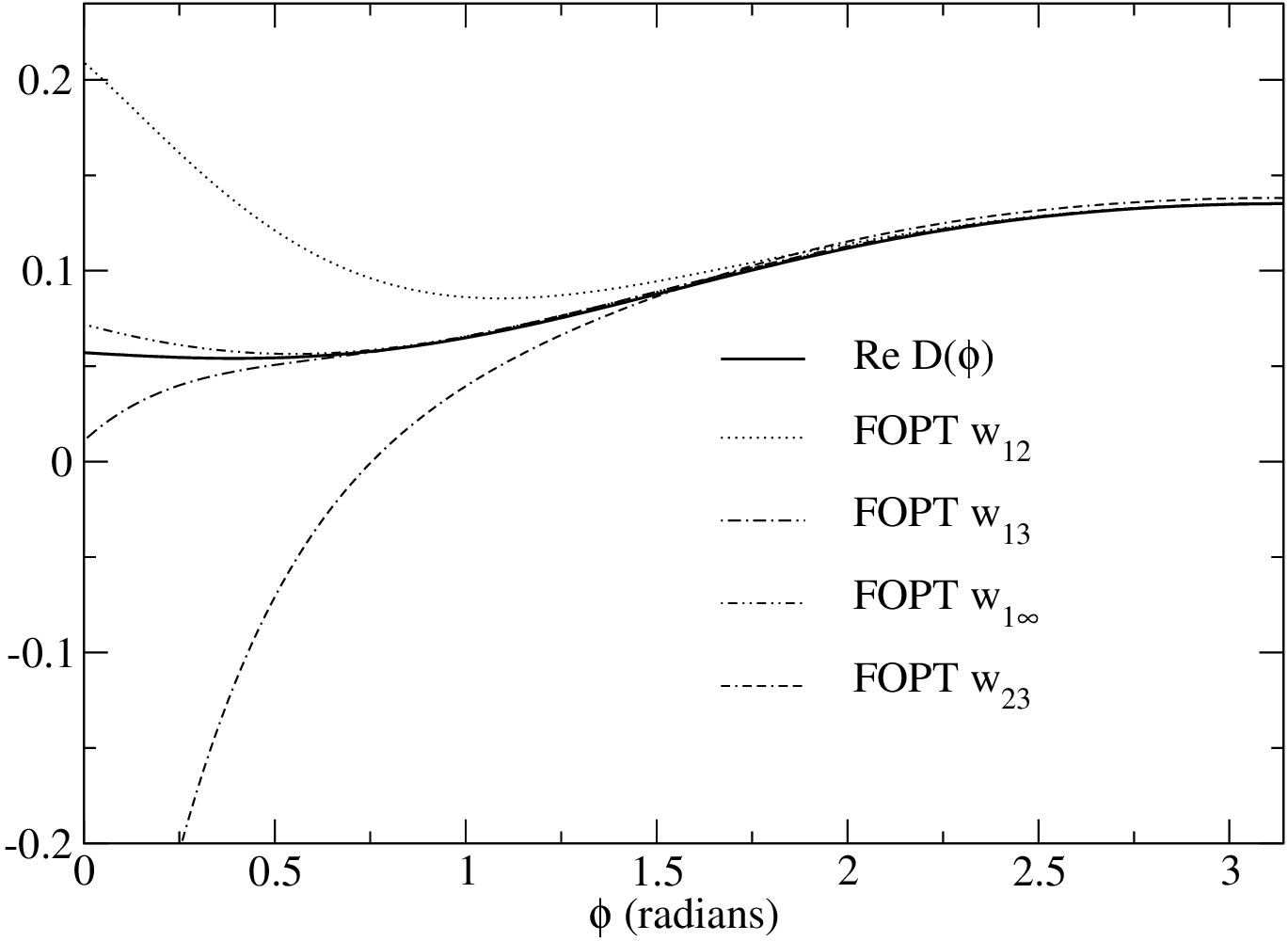}
\caption{Real part of the Adler function calculated along the circle $s=M_\tau^2 \exp(i\phi)$ with the perturbative expansions for  $N=18$. Left panel: CI expansions. Right panel: FO expansions. The exact function is represented by the solid line. The standard CIPT and FOPT exhibit big oscillations and are not shown.}\end{center}\vspace{-0.5cm}
\end{figure*}

\begin{table*}[!htb]
\caption{The quantity $\delta^{(0)}$  for a modified model with $d_2^{\rm IR}=1$, calculated for  
$\alpha_s(M_\tau^2)=0.34$  with the standard and modified  CI and FO expansions truncated at the order $N$. The rows for $N\le 5$ are identical to those in Table \ref{table:1}. Exact value $\delta^{(0)}=0.2102$.}\vspace{0.1cm}
\label{table:2}
\renewcommand{\tabcolsep}{0.5pc} 
\renewcommand{\arraystretch}{1.1} 
\begin{tabular}{llllllllllr}\hline
$N$&CIPT &FOPT& CI $w_{12}$  &  FO $w_{12}$ &CI $w_{13}$& FO $w_{13}$ & CI $w_{1\infty}$& FO $w_{1\infty}$& CI $w_{23}$ &FO $w_{23}$\\\hline 
6 & 0.2041 & 0.2318 & 0.2263 &  
0.2493 & 0.2271 & 0.2420 & 0.2284 & 0.2431 & 0.2260 & 0.2454 \\ 7 &  
0.2041 & 0.2290 & 0.2201 & 0.2628 & 0.2220 & 0.2481 & 0.2230 &  
0.2472 & 0.2174 & 0.2580 \\ 8 & 0.2023 & 0.2213 & 0.2202 & 0.2756 &  
0.2164 & 0.2595 & 0.2182 & 0.2541 & 0.2136 & 0.2734 \\ 9 & 0.2037 &  
0.2110 & 0.2175 & 0.2742 & 0.2143 & 0.2686 & 0.2154 & 0.2608 &  
0.2138 & 0.2706 \\ 10 & 0.1924 & 0.2032 & 0.2055 & 0.2709 & 0.2144 &  
0.2651 & 0.2146 & 0.2629 & 0.2115 & 0.2517 \\ 11 & 0.2124 & 0.2004 &  
0.1982 & 0.2905 & 0.2136 & 0.2504 & 0.2146 & 0.2578 & 0.2068 &  
0.2531 \\ 12 & 0.1412 & 0.2071 & 0.2007 & 0.3063 & 0.2111 & 0.2406 &  
0.2148 & 0.2468 & 0.2081 & 0.2627 \\ 13 & 0.3121 & 0.2117 & 0.2022 &  
0.2820 & 0.2086 & 0.2449 & 0.2149 & 0.2340 & 0.2060 & 0.2133 \\ 14 &  
-0.2105 & 0.2344  & 0.2001 & 0.2666 & 0.2074 & 0.2459 & 0.2146 &  
0.2239 & 0.2124 & 0.1338 \\ 15 & 1.2336 & 0.1934 & 0.2009 & 0.2865 &  
0.2079 & 0.2176 & 0.2142 & 0.2187 & 0.2087 & 0.1192 \\ 16 & -3.147 &  
0.3500 & 0.2044 & 0.2562 & 0.2091 & 0.1676 & 0.2136 & 0.2175 &  
0.2073 & 0.0930 \\ 17 & 9.948 & -0.2333 & 0.2059 & 0.1822 & 0.2102 &  
0.1355 & 0.2130 & 0.2175 & 0.2275 & -0.0415 \\ 18 & -30.94 & 2.084 &  
0.2058 & 0.1722 & 0.2107 & 0.1345 & 0.2124 & 0.2159 & 0.1617 &  
-0.1019 \\\hline 
 \end{tabular}
 \end{table*}

In the particular case of the Adler function in massless QCD, the nature of the first singularities of the Borel transform is also known: near $u=-1$, $B(u) \sim (1+u)^{-\gamma_1}$  and near $u=2$, $B(u)\sim (1-u/2)^{-\gamma_2}$. The exponents $\gamma_1$ and $\gamma_2$ are positive and are given in \cite{BeJa,Muel,BBK}. As explained in \cite{SoSu,CaFi}, this information can be exploited to further improve the convergence, by expanding the product of $B(u)$ with suitable factors that "soften" the dominant singularities.  The effect of a mild singularity,  which vanishes instead of exploding at  $u=-1$ or $u=2$, appears only at large orders in an expansion. So, at low orders we can expand in powers of variables that account only for the next branch-points of $B(u)$. In general, we consider the functions\footnote{The mapping $w_{1\infty}$, suggested in \cite{Muel1}, was applied in detail in \cite{Alta}, and the mapping  $w_{13}$ was considered in \cite{CvLe}.} 
\be\label{eq:wjk}\vspace{-0.2cm}
\wt w_{jk}(u)=\frac{\sqrt{1+u/j}-\sqrt{1-u/k}}{\sqrt{1+u/j}+\sqrt{1-u/k}},\vspace{-0.2cm}\ee
 which map the $u$-plane cut along $ u\le -j$ and $u\ge k$ onto the disc $|w_{jk}|<1$ in the plane $w_{jk}=\wt w_{jk}(u)$. We take $j\ge 1$ and $k\ge 2$, the lowest values leading to the optimal mapping (\ref{eq:w}).

As discussed in \cite{CaFi2009}, in contrast with the optimal conformal mapping, the "singularity softening" procedure is not unique. Numerically, it turns out to be convenient to define the expansion
\be \label{eq:prod} \vspace{-0.3cm}S_{jk}(u) B(u) = \sum_{n\ge 0} c_n^{jk}  (\wt w_{jk}(u))^n,\vspace{-0.3cm}
\ee
where the compensating factor  $S_{jk}(u)$ is a simple expression vanishing at $u=-1$ and $u=2$:
 \be S_{jk}(u)=\left(\!1-\frac{\wt w_{jk}(u)}{\wt w_{jk}(-1)}\!\right)^{\!\!\gamma^{(j)}_1} \!\!\left(\!1-\frac{\wt w_{jk}(u)}{\wt w_{jk}(2)}\!\right)^{\!\!\gamma^{(k)}_2}\hspace{-0.3cm}. \ee
The exponents $\gamma_1^{(j)}=\gamma_1 (1+\delta_{j1})$ and 
 $\gamma_2^{(k)}= \gamma_2 (1+\delta_{k2})$, where $\delta_{ij}$ is Kronecker's function, are taken such as to reproduce the nature of the first branch-points of $B(u)$. 

By combining the expansion (\ref{eq:prod}) with the definition (\ref{eq:pv}), we are led to the class of expansions
\be\label{eq:Djk} \wh D (s) = \sum\limits_{n} c_n^{jk} \, {\cal W}^{jk}_n(s),\vspace{-0.6cm}\ee
\be\label{eq:Wnjk} {\cal W}^{jk}_n(s)=\frac{1}{\beta_0} {\rm PV}\int\limits_0^\infty\!{\rm e}^{-\frac{u}{\beta_0 a_s(s)}} \,  \frac{(\wt w_{jk}(u))^n}{S_{jk}(u)}\, {\rm d} u.\ee

Strictly speaking, for a fixed pair ($j, k$) the expansion (\ref{eq:prod})  converges only in the disc $|w_{jk}|< \min [|\wt w_{jk}(-1)|,\, |\wt w_{jk}(2)|]$. In particular, for $j=1$ the expansions diverge for $|u|>2$ if $k>2$ (for $k=2$ the expansion converges outside the disc $|u|=2$, but not on the real axis \cite{CaFi}). However, the expansion (\ref{eq:prod})  enters the Laplace-Borel integral (\ref{eq:pv}) where, especially for small couplings $a_s(s)$, the contribution of high values of $u$   is suppressed, reducing the effect of the lack of convergence.   On the other hand, if $j>1$ the expansion (\ref{eq:prod})  does not converge for $u>1$, and the effect of this divergence is expected to become visible earlier. 
By inserting into (\ref{eq:Wnjk}) the coupling $a_s(s)$ calculated by solving the renormalization group equation for $s$ along the circle defined in the integral (\ref{eq:delta0}), we obtain  the "countour-improved"
(CI) version of the new expansions. The "fixed-order" (FO) version  can be obtained in a straightforward way \cite{CaFi2009}. The  expansion functions ${\cal W}^{12}_n$ were investigated in \cite{CaFi2009}. In what follows we shall investigate also the expansion in terms of ${\cal W}^{13}_n$, ${\cal W}^{1\infty}_n$ and ${\cal W}^{23}_n$.

\section{MODELS}\label{sec:models}

For testing the convergence of the various expansions we consider a class of models of the type proposed in \cite{BeJa}, which parametrize the Borel function $B(u)$ in terms of a few UV and IR singularities and recover the Adler function by means of (\ref{eq:pv}). The free parameters are  fixed by  reproducing the known values of the coefficients $K_n$ for $n\le 4$. The specific model proposed in \cite{BeJa} (which uses as input also $K_5=283$), leads to a rather large  residue, $d_2^{\rm IR}=3.13$, for the first IR renormalon. To avoid any bias related to this large contribution, we investigated also  alternative models having a smaller residue, $d_2^{\rm IR}=1$ (details and more results  will be given in \cite{CaFinew}). 

For illustration we give in  Table \ref{table:1} the values of the integral $\delta^{(0)}$ defined in (\ref{eq:delta0}) for the model \cite{BeJa}, calculated  with the standard and the modified  CI and FO expansions, as a function of the perturbative order $N$. As discussed in  \cite{BeJa} the standard CIPT gives values systematically lower than the true result, while the standard FOPT gives a better approximation.  The new CI expansions approach the exact value for increasing $N$ (deviations appear only for the mapping $w_{23}$, for the reasons discussed above). The new FO expansions give in general a less accurate description, but with the mappings $w_{13}$ and $w_{1\infty}$ convergent approximations are obtained also in the FO case.

The numbers in Table \ref{table:1} can be understood from Figs. 1 and 2,  which show the real part of the Adler function along the circle $|s|=M_\tau^2 \exp(i\phi)$, calculated with the CI and FO expansions truncated at $N=5$ and $N=18$, respectively.  The CI expansions  give a precise approximation of $\wh D(s)$ along the whole circle (only  the mapping  $w_{23}$ shows signs of divergence for $N=18$, as expected). In the FO case the description gradually deteriorates near the timelike axis ($\phi=0$) due to the poor convergence of the expansion of $\alpha_s(s)$ in powers of  $\alpha_s(M_\tau^2)$ in this region \cite{DiPi}. Note that the coupling $a_s(s)$ is calculated in the same way in the exact representation (\ref{eq:pv}) and the CI expansion functions (\ref{eq:Wnjk}), as the exact solution of the RG equation in terms of $\alpha_s(M_\tau^2)$, while in the FO version $a_s(s)$ is expanded in powers of $\alpha_s(M_\tau^2)$. Of course, if the CI expansions become very precise at high $N$,  the FO expansions are expected to converge too (this is indeed visible in the right panel of Fig. 2 for the mappings $w_{13}$ and $w_{1\infty}$).

 In Table \ref{table:2} we present similar results for an alternative model. By construction the first five rows in Tables \ref{table:1} and \ref{table:2} are the same. The CI expansions based on the mappings  $w_{12}$, $w_{13}$ and $w_{1\infty}$ approach at large $N$ the exact value also in this case.  In the FO case the description is less precise and,  for the values of $N$ considered, only the expansion based on  $w_{1\infty}$  exhibit convergence.

\begin{table}[!htb]\begin{center}\vspace{-0.5cm}\caption{}\vspace{-0.3cm}\label{table:alpha}
\renewcommand{\tabcolsep}{1.4pc} 
\renewcommand{\arraystretch}{1.2} 
\begin{tabular}{c c}\hline
Expansion functions & $ \alpha_s(M_\tau^2)$\\\hline
 ${\cal W}_n^{12} $ & 0.3198 (0.3200)\\
  ${\cal W}_n^{13}$  & 0.3212 (0.3212)\\
${\cal W}_n^{1\infty}$ & 0.3187 (0.3187)\\
 ${\cal W}_n^{23}$ & 0.3197 (0.3198) \\\hline
\end{tabular}\end{center}\vspace{-0.4cm}
\end{table}

\vspace{-0.5cm}
\section{DETERMINATION of $\alpha_s$}\label{sec:disc}

We apply now the new CI expansions for a determination of  $\alpha_s(M_\tau^2)$. Note that we do not use the models discussed in the previous Section, but only the first four $K_n$ known in the $\overline{\rm MS}$ scheme \cite{BCK08}, and an estimate of $K_5$.  The results are given in Table \ref{table:alpha}. The first numbers are obtained with the phenomenological value $\delta^{(0)}=0.2402$ and  $K_5=283$  \cite{BeJa}, those in parantheses with $\delta^{(0)}=0.2038$ and $K_5=275$  \cite{Pich2010}. With the same input the standard CIPT gives $\alpha_s(M_\tau^2)= 0.3425\,  (0.3421)$.

The values in Table \ref{table:alpha}  are very close to each other, so the uncertainty related to the choice of the expansion is very small. Taking the average and adding the error according to the analysis in \cite{CaFi2009} leads to 
\be\label{eq:aver}
\alpha_s(M_\tau^2)=  0.320\pm 0.011. 
\ee 
The average, consistent with the determination based on standard FOPT \cite{BeJa}, is smaller by about 0.02 than the predictions \cite{Davier2008,Pich2010} of the standard CIPT. Sorting out this discrepancy is important for the precision tests of QCD in $\tau$ decays.

\subsection*{Acknowledgements} IC acknowledges support from CNCSIS in the program Idei, Contract Nr.464/2009.


\vspace{0.01cm}
\begin{thebibliography}{9}
\bibitem{Bethke} S. Bethke,  Eur. Phys. J.C{\bf 64} (2009) 689.
\bibitem{BCK08} P.A. Baikov, K.G. Chetyrkin and J.H. K{\"u}hn,  Phys. Rev. Lett. {\bf 101} (2008) 012002.
\bibitem{DiPi}  F. Le Diberder and A. Pich, Phys. Lett. B{\bf 286} (1992) 147. 
\bibitem{Davier2008} M. Davier, S. Descotes-Genon, A. Hocker, B. Malaescu and Z. Zhang,  Eur. Phys. J. C{\bf 56} (2008) 305.
\bibitem{BeJa} M. Beneke and M. Jamin, JHEP {\bf 09} (2008) 044. 
\bibitem{Pich2010} A. Pich,  Acta Phys.Polon.Supp. 3 (2010) 165. 
\bibitem{CaFi2009} I. Caprini and J. Fischer,   Eur. Phys. J.C{\bf 64} (2009) 35.
\bibitem{CaFi} I. Caprini and J. Fischer, Phys. Rev.  D{\bf 60} (1999) 054014; id.  D{\bf 62} (2000) 054007; Eur. Phys. J. C{\bf 24} (2002) 127.
\bibitem{Dyson} F.J. Dyson, Phys. Rev. {\bf 85} (1952) 631.
\bibitem{CaFiVr} I. Caprini, J. Fischer and I. Vrko\v{c},  J. Phys. A  Math. Theor. {\bf 42} (2009)  395403. 
\bibitem{CaNe} I. Caprini and M. Neubert, JHEP {\bf 03} (1999) 007. 
\bibitem{CiFi} S. Ciulli and J. Fischer, Nucl. Phys. {\bf 24} (1961) 465.
\bibitem{CaFinew} I. Caprini and J. Fischer, to be published.
\bibitem{Muel} A. Mueller, Nucl.Phys. B{\bf 250} (1985) 327.
\bibitem{BBK} M. Beneke, V.M. Braun and N. Kivel, Phys. Lett. B{\bf 404} (1997) 315.
\bibitem{SoSu} D.E. Soper and L.R. Surguladze, Phys. Rev. D{\bf 54} (1996) 4566.
\bibitem{Muel1} A.H. Mueller, in {\em QCD - Twenty Years Later}, Aachen 1992, 
edited by P. Zerwas and H. A. Kastrup (World Scientific, Singapore, 1992).
\bibitem{Alta}   G. Altarelli, P. Nason and G. Ridolfi, Z. Phys. C{\bf 68} (1995) 257.
\bibitem{CvLe} G. Cveti\v c and T. Lee, Phys. Rev D{\bf 54} (2001) 014030.
 
\end{thebibliography}
\end{document}